\documentclass[aps,prc,twocolumn,floatfix,nofootinbib,preprintnumbers,superscriptaddress,longbibliography]{revtex4-1}

\usepackage{epsfig}
\usepackage{graphicx}
\usepackage{stmaryrd}
\usepackage{amssymb,tabularx,dcolumn}
\usepackage{supertabular,ltxtable}
\usepackage{longtable}
\usepackage{amsmath}
\usepackage{float}
\usepackage{color}
\usepackage{bm}
\usepackage{CJKutf8}
\usepackage{hyperref}
\usepackage{mathtools}
\usepackage{multirow}
\usepackage{makecell}
\usepackage[utf8]{inputenc}
\usepackage{xcolor}
\usepackage{soul}

\makeatletter
\def\@bibdataout@aps{%
\immediate\write\@bibdataout{%
@CONTROL{%
apsrev41Control%
\longbibliography@sw{%
    ,author="08",editor="1",pages="1",title="0",year="1"%
    }{%
    ,author="08",editor="1",pages="1",title="",year="1"%
    }%
  }%
}%
\if@filesw \immediate \write \@auxout {\string \citation {apsrev41Control}}\fi 
}
\makeatother

\begin{document}

\begin{CJK*}{UTF8}{gbsn}

\title{Landscape of pear-shaped even-even nuclei}

\author{Yuchen Cao (曹宇晨)}
\affiliation{National Superconducting Cyclotron Laboratory, Michigan State University, East Lansing, Michigan 48824, USA}
\affiliation{Department of Physics and Astronomy, Michigan State University, East Lansing, Michigan 48824, USA}
\author{S. E. Agbemava}
\affiliation{Department of Physics and Astronomy, Mississippi State University, Mississippi 39762, USA}
\author{A. V. Afanasjev}
\affiliation{Department of Physics and Astronomy, Mississippi State University, Mississippi 39762, USA}
\author{W. Nazarewicz}
\affiliation{Facility for Rare Isotope Beams, Michigan State University, East Lansing, Michigan 48824, USA}
\affiliation{Department of Physics and Astronomy, Michigan State University, East Lansing, Michigan 48824, USA}
\author{E. Olsen}
\affiliation{Institut d'Astronomie et d'Astrophysique, 
Universit{\'e} Libre de Bruxelles, 1050 Brussels, Belgium}

\date{\today}

\begin{abstract}
\begin{description}
\item[Background]
The phenomenon of reflection-asymmetric nuclear shapes is relevant to
nuclear stability, nuclear spectroscopy, nuclear decays and fission, and  the search   for new physics beyond the standard model. Global surveys of ground-state octupole deformation, performed with a limited number of models, suggest that the number of pear-shaped isotopes is fairly limited across the nuclear landscape.

\item[Purpose]
We carry out global analysis of ground-state octupole deformations for  particle-bound even-even nuclei with $Z \leq 110$ and $N \leq 210$ using nuclear density functional theory (DFT) with several non-relativistic and covariant energy density functionals. In this way, we can identify the best candidates for reflection-asymmetric shapes.

\item[Methods]
The calculations are performed in the frameworks of axial reflection-asymmetric Hartree-Fock-Bogoliubov theory and relativistic Hartree-Bogoliubov theory using 
DFT solvers employing harmonic oscillator basis expansion. We consider five Skyrme and four covariant energy density functionals.

\item[Results]
We predict several regions of ground-state octupole deformation. In addition to the ``traditional" regions of neutron-deficient actinide nuclei around $^{224}$Ra and 
neutron-rich lanthanides around $^{146}$Ba, we identified vast regions of reflection-asymmetric shapes in very neutron-rich nuclei around $^{200}$Gd and $^{288}$Pu, as well as in several nuclei around  $^{112}$Ba. Our analysis suggests  several promising candidates with stable ground-state octupole deformation, primarily in the neutron-deficient actinide region, that can be reached experimentally.
Detailed comparison between Skyrme and covariant models is performed.

\item[Conclusions]
Octupole shapes predicted in this study are consistent with the current experimental information. This work can serve as a starting point of a systematic  search for
parity doublets in odd-mass and odd-odd nuclei, which can be of interest in the context of 
new physics searches.
\end{description}
\end{abstract}

\maketitle
\end{CJK*}

\section{Introduction}\label{intro}

The majority of atomic nuclei have reflection-symmetric ground-states (g.s.),  and exhibit either spherical
or ellipsoidal (prolate or oblate) shapes.
In rare cases, however, the nucleus can  spontaneously break its intrinsic reflection symmetry as a result of a nuclear Jahn-Teller effect \cite{JahnTeller,PGOtten,Nazarewicz94,Frauendorf01}
and acquire non-zero octupole moments associated with  pear-like shapes \cite{Strutinsky,Lee1957,Johansson1961} (see Refs. \cite{Butler1996,Butler2016} for comprehensive reviews).

Early systematic calculations of octupole shapes were carried out with a macroscopic-microscopic (MM) approach based on the shell correction method
\cite{Gyurkovich1981,Witek1984,Moller1995} (see also Ref.~\cite{Moller2008} for an update). Those were followed by self-consistent studies within nuclear DFT with Gogny \cite{Robledo2011,Robledo2012,Warda2012,Robledo2015}, BCP \cite{Robledo2010,Robledo2012}, Skyrme \cite{Erler2012a,Ebata2017}, and covariant \cite{Nomura2014,Xu2017,Agbemava2016,Agbemava2017} energy density functionals (EDFs).

Except for the global surveys \cite{Moller2008,Robledo2011,Agbemava2016,Xia2017,Ebata2017}, the majority of the previous DFT studies were focused on three specific regions of  octupole collectivity: neutron-deficient  actinides, neutron-rich lanthanides, and neutron-rich heavy and superheavy nuclei that are important for the modeling of heavy-element nucleosynthesis. Consequently, to better understand systematic trends of octupole instability throughout the nuclear landscape, it is helpful to carry out  additional  global inter-model comparisons: that is the main objective of this work.

Additionally, the results of this study can provide robust candidates for atomic parity and time-reversal  violation searches \cite{Haxton1983,Haxton1997, Chupp2019}. Of particular interest is the atomic electric dipole moment (EDM) \cite{Chupp2019}. 
The nuclear quantity behind the atomic EDM is the Schiff moment~\cite{Schiff1963}, which can be enhanced by the presence of nuclear octupole deformation~\cite{Auerbach1996,Spevak1997,Engel2000,Engel2003,Flambaum2019}. In particular, the recent study \cite{Jacek2018} has demonstrated a correlation between the nuclear octupole deformation and the Schiff moment.  Thus it is believed that the best candidates for atomic EDM measurements, such as $^{225}$Ra, are nuclei with octupole shapes. A strong motivation of this work is to produce data for  systematic Schiff moment calculations in  odd-mass and odd-odd nuclei in the vicinity of the most robust   octupole-deformed even-even candidates.

This paper is organized as follows. Section~\ref{theory} describes the theoretical frameworks  used. The results of our global calculations and an analysis of trends are presented in Sec.~\ref{global}. 
 The discussion of local regions  of octupole-deformed nuclei is done in Sec.~\ref{local}. Finally, Sec.~\ref{summary}  contains a summary and conclusions.

\section{THEORETICAL FRAMEWORK}\label{theory}

Our calculations are performed in the framework of nuclear DFT \cite{Bender2003}. They are restricted to axial shapes, as triaxiality affects the ground states of only a limited set of nuclei which do not overlap with regions of octupole deformation \cite{Moller2008,Ebata2017}.  Moreover, the Skyrme DFT calculations of Ref.~\cite{Ebata2017}, which allow for non-axial reflection asymmetric shapes, show that octupole deformed lanthanides and actinides are axially symmetric. 

A comment on our model selection is in order. To be useful in an analysis such as ours, a model must meet several stringent criteria.  First, since it is meant to be used for extrapolations into  unknown regions of the nuclear landscape, the selected model should be based on controlled many-body formalism employing quantified input (here: the energy density functional). Second, the underlying theoretical framework should be capable of reproducing basic nuclear properties  (such as masses, radii, shell structure, and deformations). Finally, the model should be globally applicable throughout the nuclear chart.  The models employed in this study  are based on energy density functionals that were tested globally and which are consistent with bulk ground-state data. By using such validated models, we feel comfortable making  predictions for new observables, such as octupole moments.
The details pertaining to  
axial reflection-asymmetric Hartree-Fock-Bogoliubov (HFB) calculations (Sec.~\ref{SHFB}) and relativistic Hartree-Bogoliubov (RHB) calculations  (Sec.~\ref{RHB}) can be found in Refs.~\cite{Erler2012} and \cite{Agbemava2016}, respectively. 

We studied particle-bound  even-even nuclei with 
 $Z \leq 110$ and $N \leq 210$.   For nuclei with  $Z \geq 112$, Coulomb frustration effects result in  
 exotic topologies of nucleonic densities  such as bubbles and tori~\cite{Afa05,Bastian2017,Witek2018,AAG.18,Giuliani2019,AATG.19}, 
  which give rise to instabilities of potential energy surfaces due to configuration changes. 
 
The axial shape deformations $\beta_\lambda$ are defined through  
multipole moments $Q_{\lambda 0}$:
 \begin{equation}
\begin{aligned} \beta_{2} &=Q_{20} /\left(\sqrt{\frac{16 \pi}{5}} \frac{3}{4 \pi} A R_{0}^{2}\right), \\ \beta_{3} &=Q_{30} /\left(\sqrt{\frac{16 \pi}{7}}\frac{3}{4 \pi} A R_{0}^{3}\right), \end{aligned}
\label{def-def}
\end{equation}
where $R_0 = 1.2A^{1/3}$ and 
\begin{equation}
\begin{array}{l}{{Q}_{20}= \langle 2z^{2}-x^{2}-y^{2} \rangle}, \\[10pt] {{Q}_{30}= \langle z\left(2 z^{2}-3 x^{2}-3 y^{2}\right)\rangle}.\end{array}
\end{equation}
Total/proton multipole moments are used in the HFB/RHB calculations, respectively. This difference is not critical 
since proton and neutron deformations are very similar in the range of nuclei considered. Note also that
in the  DFT calculations, all multipole moments (in our case, $Q_{20}$, $Q_{30}$, $Q_{40}$, $Q_{50}$, $Q_{60}$, ...)
that correspond to the energy minimum are obtained from the self-consistent particle density and the respective deformations ($\beta_2$, $\beta_3$, $\beta_4$, 
$\beta_5$, $\beta_6$, ...) are  computed from these moments. This is contrary to the
MM approach in which the energy minimization is usually performed in  a multidimensional
space of  selected deformations.

The magnitude of octupole deformation alone is insufficient in determining whether robust octupole 
deformation is present since it does not provide any information on the softness of
 the potential energy surface in the octupole direction. 
To address this  issue, we  also look at the gain in binding energy $\Delta E_{\rm oct}$ due to octupole deformation:
\begin{equation}
\Delta E_{\mathrm{oct}}=E^{\mathrm{a}}\left(\beta_{2}, \beta_{3}\right)-E^{\mathrm{s}}\left(\beta_{2}^{\prime}, \beta_{3}^{\prime}=0\right),
\label{energy-gain}
\end{equation}
where $E^{\rm a}$ is the absolute binding energy obtained in  reflection-asymmetric calculations, and $E^{\rm s}$ is the  binding energy minimum from reflection-symmetric calculations. These two minima do not necessarily have the same quadrupole deformation. $\Delta E_{\mathrm{oct}}$  is also an indicator of the stability of octupole-deformed shapes, where large values are typical for potential energy surfaces (PESs) with well-pronounced octupole minima; for such systems, the concept of  static octupole deformation is better justified \cite{Butler1996}. Conversely, small $\Delta E_{\mathrm{oct}}$
 values are characteristic of octupole-soft PESs typical of octupole vibrations; for such systems beyond-mean-field effects can play an important role \cite{Egido1991,Robledo2010,Robledo2012,Robledo2015,Robledo2016,Yao2015,Yao2016,Xia2017}.

\subsection{Skyrme-Hartree-Fock-Bogoliubov calculations}\label{SHFB}

In our study, we consider five Skyrme energy density functionals (SEDFs): UNEDF0 \cite{UNEDF0}, UNEDF1 \cite{UNEDF1}, UNEDF2 \cite{UNEDF2}, SLy4 \cite{SLy4}, and SV-min~\cite{SV-min}.  These SEDFs are described by means of 12$\sim$14  coupling constants. The root-mean-square (rms) error of binding energy of these SEDFs, compared to the experimental mass dataset AME2016~\cite{AME2016} ranges from 1.7 MeV (UNEDF0) to 5.3 MeV (SLy4). In the pairing channel, we took the
mixed-type density-dependent delta interaction~\cite{dobaczewski2002}
with Lipkin-Nogami approximate particle-number projection as in
Ref.~\cite{Stoitsov2003}.  The pairing strengths for SLy4 and SV-min were assumed to be $-258.2$\,MeV and $-214.28$\,MeV, respectively, assuming the same value for neutrons and protons.

The calculations were performed using the parallel DFT solver HFBTHO (v3.00) \cite{HFBTHO300} that solves the HFB equation in the cylindrical deformed harmonic oscillator basis. 
We utilized the ``kick-off" mode~\cite{HFBTHO300}, whereby the multipole moments are constrained in the initial ``kick-off" stage  and subsequently released when certain criteria are met. Dynamic MPI scheduling was implemented to further reduce computational cost in large-scale mass-table calculations.

The effectiveness and efficiency of the ``kick-off" mode has been thoroughly tested and benchmarked with  PES calculations of more than a hundred nuclei in various mass regions. A cylindrical harmonic oscillator basis of $N = 20$ major oscillator shells was used; this was tested to be equivalent (within a reasonable accuracy) in the prediction of g.s. masses compared to using larger shell numbers. Computational savings by using ``kick-off" mode and dynamic MPI scheduling,  compared with PES calculations under static MPI scheduling, were found to be very significant.

\begin{figure*}[!htb]
\includegraphics[width=\linewidth,keepaspectratio]{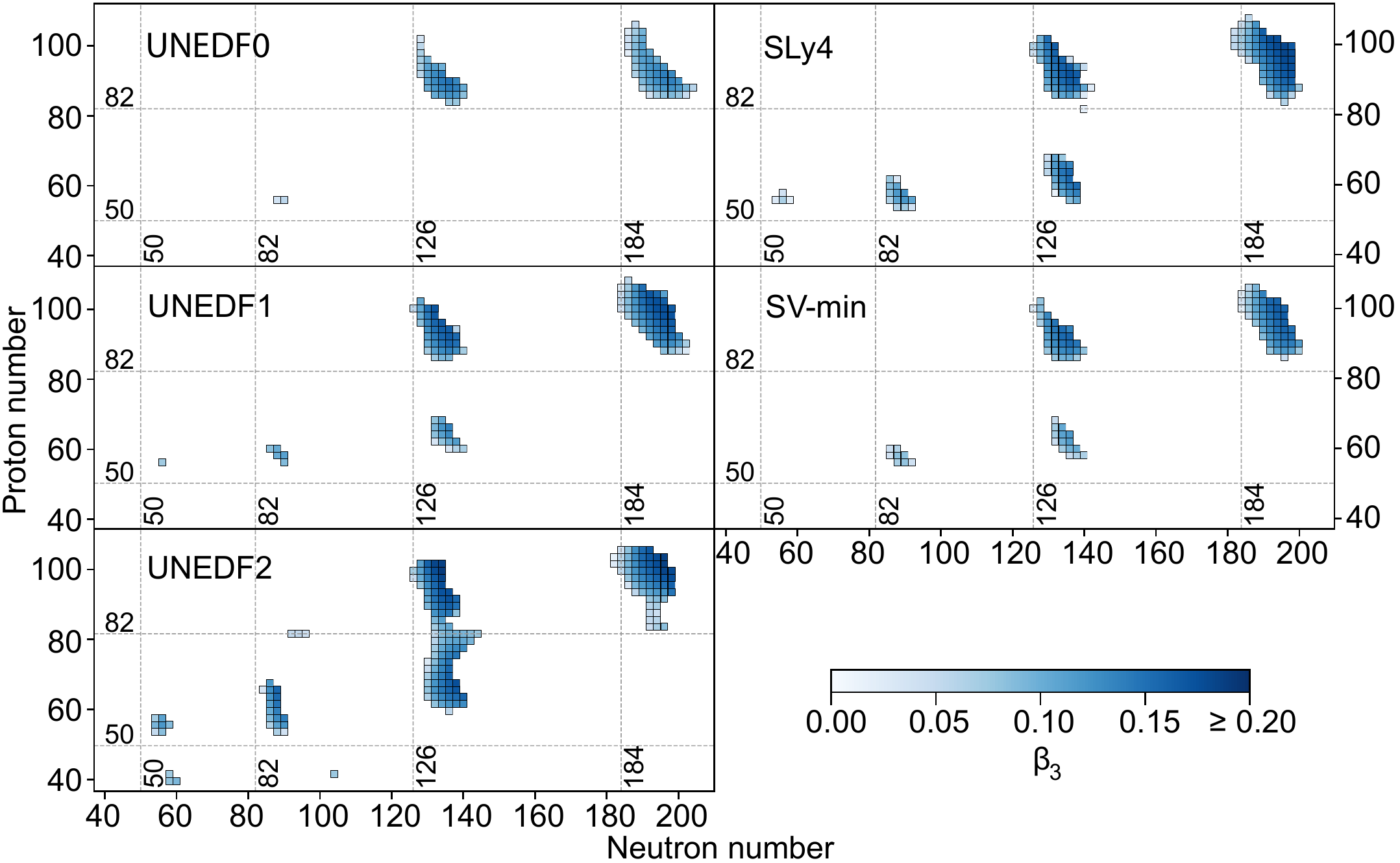}
\caption{\label{beta3_plot} Total g.s. octupole deformations $\beta_3$ of even-even nuclei in the $(Z,N)$ plane predicted with the SEDFs UNEDF0, UNEDF1, UNEDF2, SLy4, and SV-min. The magic numbers are indicated by dashed lines.
}
\end{figure*}

\begin{figure*}[!htb]
\includegraphics[width=\linewidth,keepaspectratio]{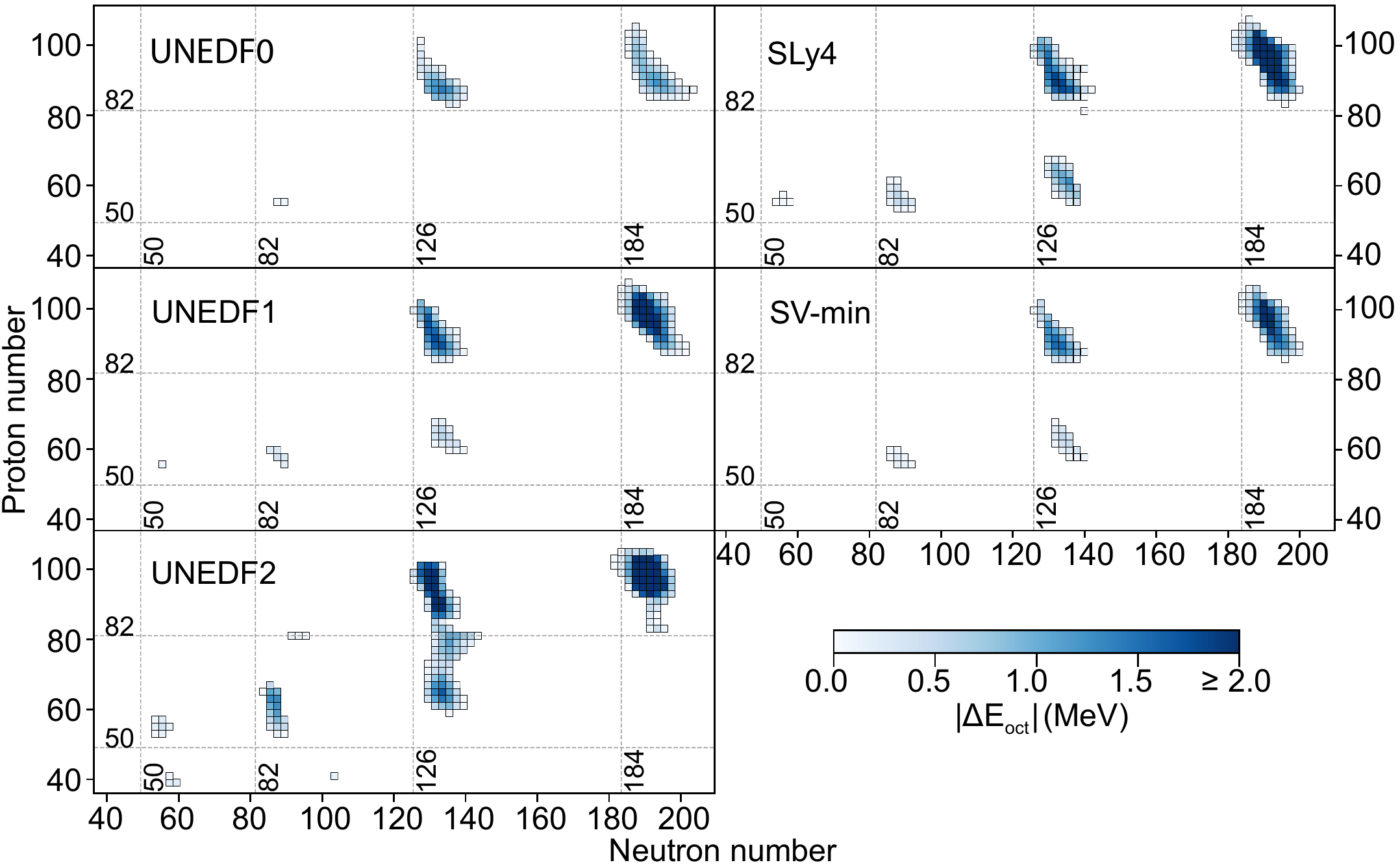}
\caption{\label{Eoct}  Similar to Fig.~\ref{beta3_plot} but for the octupole deformation energy  $\Delta E_{\rm oct}$.
}
\end{figure*}

\subsection{Relativistic Hartree-Bogoliubov calculations}\label{RHB}

In the covariant DFT, the nucleus is considered as a system of 
$A$ nucleons which interact via the exchange of different mesons \cite{VALR.05}. A global search
for octupole deformed nuclei has been performed using four covariant energy 
density functionals (CEDFs): DD-ME2 \cite{DD-ME2}, NL3* \cite{NL3*}, PC-PK1 
\cite{PC-PK1} and DD-PC1 \cite{DD-PC1}.  These CEDFs represent 
three major classes of covariant DFT models: the non-linear meson-nucleon 
coupling model (represented by NL3*), the density-dependent meson exchange 
model (represented by the DD-ME2) and the point coupling  
model (represented by  DD-PC1 and PC-PK1).  These functionals typically contain 6 to 9 
parameters which  are fitted to experimental data on finite nuclei and nuclear matter properties \cite{AARR.14}. We used
separable pairing of finite range  \cite{TMR.09} with the strength defined as in
 Ref.~\cite{AARR.14}.
As compared with the experimental AME2012 dataset,  the rms error of binding energy of 
these CEDFs ranges from 2.15 MeV (DD-PC1) to 3.0 MeV (NL3*) \cite{AARR.14}.

    The reflection-asymmetric RHB calculations are carried out using a parallel version
of the computer code developed in Ref.~\cite{Agbemava2016}, formulated in an axially deformed harmonic oscillator
basis.
In the present paper, additional calculations  to those presented in Refs.\ 
\cite{Agbemava2016,Agbemava2017} have been performed to cover the same range 
of nuclei as in the Skyrme HFB calculations. The procedure, similar to the
``kick-off" procedure employed in the HFBTHO calculations, is also used in the
RHB calculations.  However, in this case the set of initial Woods-Saxon pear-like
densities defined by the basis deformations
are used at the initial step of the calculations to push  the convergence to the octupole
deformed minimum.

\section{Global survey}\label{global}

The  g.s. octupole deformations $\beta_3$ obtained in our HFB calculations are displayed   in Fig.~\ref{beta3_plot}. (For RHB results, see Refs.~\cite{Agbemava2016,Agbemava2017}.) There is a good inter-model consistency, with  
 large octupole deformations predicted around $^{146}$Ba (neutron-rich lanthanides), $^{200}$Gd (very neutron-rich lanthanides), $^{224}$Ra (neutron-deficient actinides), and $^{288}$Pu (neutron-rich actinides), i.e., in the regions of strong octupole collectivity defined by the presence of close-lying proton and neutron shells with $\Delta\ell=\Delta j=3$~\cite{Butler1996}. This finding is consistent  with previous global studies~\cite{Moller2008,Robledo2011,Agbemava2016,Xia2017,Ebata2017}.

In each region of octupole-deformed nuclei, the magnitude of octupole deformation  increases with   the number of valence nucleons. All five SEDFs predict neutron-deficient and neutron-rich actinides to exhibit strong octupole deformations, while predictions in the lanthanide region are less uniform regarding which nuclei are deformed and how deformed they are. In general, UNEDF2 and SLy4 predict the largest number of octupole-deformed nuclei and also the larger  values of $\beta_3$.
In both models, proton-rich nuclei around $^{112}$Ba  are expected to be reflection-asymmetric.
The functional  UNEDF0 predicts the least amount of octupole-deformed nuclei and smaller $\beta_3$  deformations overall.

The octupole deformation energies  $\Delta E_{\rm oct}$  predicted in our HFB calculations are shown in Fig.~\ref{Eoct}. (For RHB results, see Refs.~\cite{Agbemava2016,Agbemava2017}.) We can see that  lanthanide nuclei have appreciably  smaller $\Delta E_{\rm oct}$ values as compared to the actinides in spite of similar octupole deformations. This indicates that most of the reflection-asymmetric lanthanide nuclei are  predicted to  have very soft PESs in the octupole direction, regardless of the equilibrium value of $\beta_3$. 

\begin{figure}[!htb]
\includegraphics[width=0.8\linewidth]{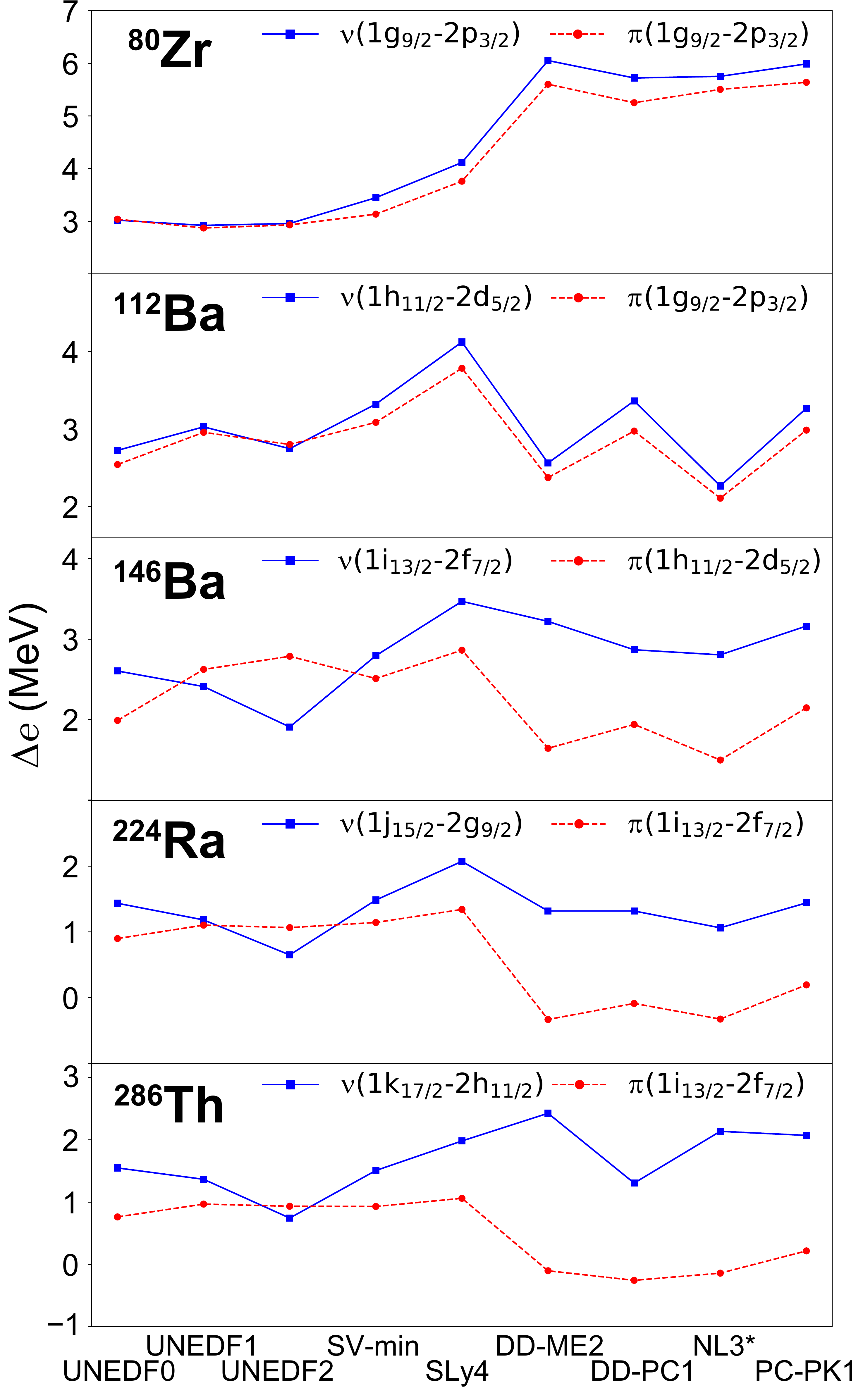}
\caption{\label{sp_split} Single-particle energy splitting $\Delta e$ between 
the unusual-parity intruder shell $(\ell,j)$ and 
the normal-parity shell $(\ell-3, j-3)$ for five nuclei representing  different regions  of octupole instability. The s.p. canonical states were obtained from spherical HFB/RHB calculations.
The neutron (proton) splittings are indicated by the solid (dashed) lines.
}
\end{figure}

Microscopically, octupole deformations can be traced back to close-lying
pairs of single-particle (s.p.) shells
coupled by the octupole interaction~\cite{Butler1996}. Each pair consists of the unusual-parity intruder shell $(\ell,j)$ and 
the normal-parity shell $(\ell-3, j-3)$. Consequently,
the regions of nuclei with strong
octupole correlations correspond to particle numbers
near 34 ($g_{9/2}\leftrightarrow p_{3/2}$ coupling), 
56 ($h_{11/2}\leftrightarrow  d_{5/2}$), 88 ($i_{13/2}\leftrightarrow f_{7/2}$), 134 ($j_{15/2} \leftrightarrow  g_{9/2}$), and 196 ($k_{17/2} \leftrightarrow  h_{11/2}$).

Figure~\ref{sp_split} shows the energy splitting 
\begin{equation}\label{Deoct}
\Delta e = e(\ell,j)-e(\ell-3,j-3),
\end{equation}
between s.p. canonical shells  obtained from spherical HFB/RHB calculations.
In general, there is a systematic decrease of $\Delta e$ with mass, which  -- together with the increased degeneracy of s.p. orbits (and matrix elements of the octupole coupling)  -- results in enhanced  octupole correlations in heavy nuclei.  However, while this general trend is robust, the magnitude of $\Delta e$  is not a good indicator of octupole correlations when comparing different models.
Indeed, when comparing different models one also needs to consider other factors related to each model's structure. For instance, the isoscalar effective mass of SLy4 is close to 0.7, which effectively increases the s.p. splitting as compared to UNEDF models (which have effective mass close to one). As a result, although in most cases SLy4 has larger $\Delta e$  than UNEDF1, it predicts more octupole-deformed nuclei and larger $\Delta E_{\rm oct}$ values.
When  comparing predictions of the UNEDF family of SEDFs, the UNEDF2 parametrization constrained to the spin-orbit splittings in several nuclei yields the lowest values of $\Delta e$ for neutrons
and  predicts the strongest octupole correlations, see Figs.~\ref{beta3_plot} and \ref{Eoct}. Still the relation between the size of $\Delta e$  and the appearance of octupole deformations is very indirect: while the very appearance of $\Delta j=\Delta \ell =3$  doublets with low $|\Delta e|$ creates favorble conditions for reflection-asymmetric shapes in a given region, this does not tell anything about the magnitude of the symmetry breaking.

\begin{figure*}[!htb]
\includegraphics[width=0.8\linewidth]{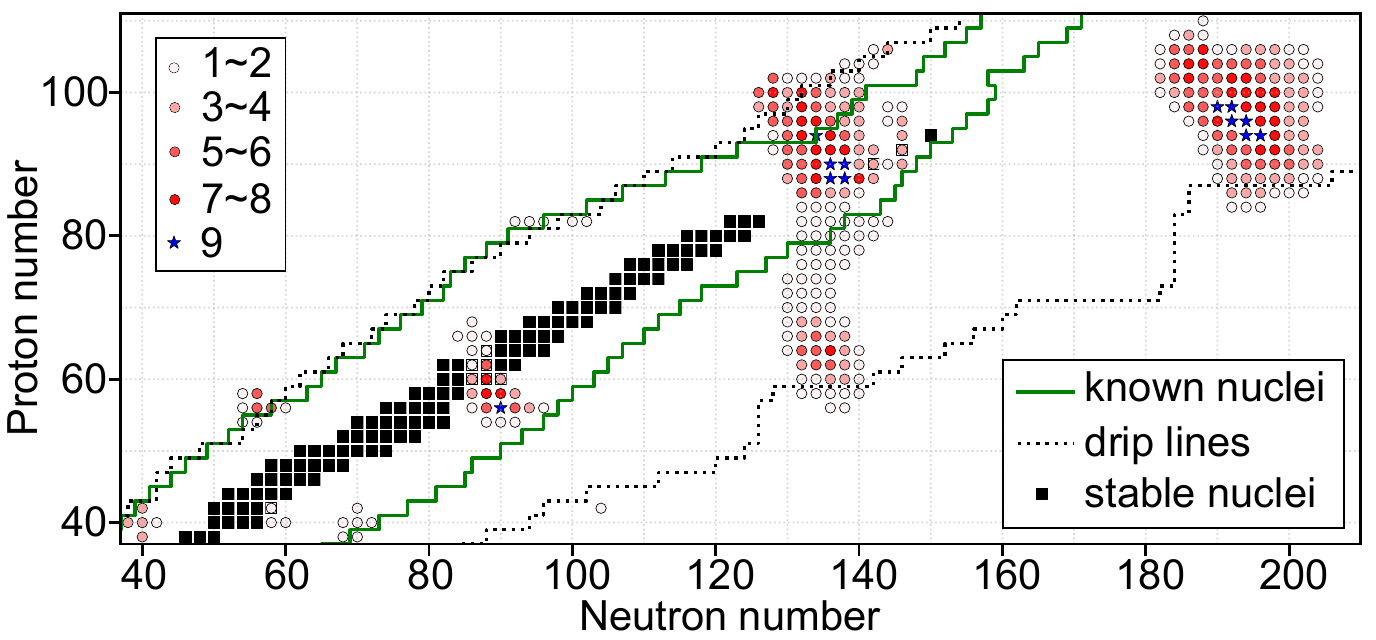}
\caption{\label{multiplicity} The landscape of g.s. octupole deformations in even-even nuclei. Circles and stars represent nuclei predicted to have nonzero octupole deformations. 
The model multiplicity $m(Z,N)$ is indicated by the legend.
 The boundary of known (i.e., experimentally discovered) nuclei is marked by the solid  line. For simplicity, this boundary is defined by the lightest and heaviest isotopes discovered for a given element. The  average two-nucleon drip lines from Bayesian machine learning studies~\cite{landscape2020,proton_emitter2020} are marked  by  dotted lines. Primordial nuclides~\cite{NNDC} are indicated by  squares.
}
\end{figure*}

In an effort to obtain a more robust picture of octupole deformations, we combined the octupole predictions from the five SEDFs and four CEDFs in Fig.~\ref{multiplicity}. 
We define the model multiplicity $m(Z, N ) = k$ if a nucleus $(Z,N)$ is predicted by $k$ models $(k = 1,\dots 9)$ to have a nonzero octupole deformation.
Nuclei predicted by all nine EDFs as octupole-deformed (i.e., $m=9$) are shown by  stars. These are:  $^{146}$Ba, $^{224,226}$Ra, $^{226,228}$Th, and $^{228}$Pu in the regions experimentally accessible, and in the very neutron-rich  actinides: $^{288,290}$Pu, $^{288,290}$Cm, and $^{288,290}$Cf. The supplemental Table \cite{SM} contains predicted values of  $\Delta E_{\rm oct}$ and $\beta_3$ values of nuclei with multiplicity $\geq 6$ as well as proton quadrupole and octupole moments in nuclei with $\beta_3 \geq 0.01$.

\begin{figure*}[!htb]
\includegraphics[width=0.8\linewidth]{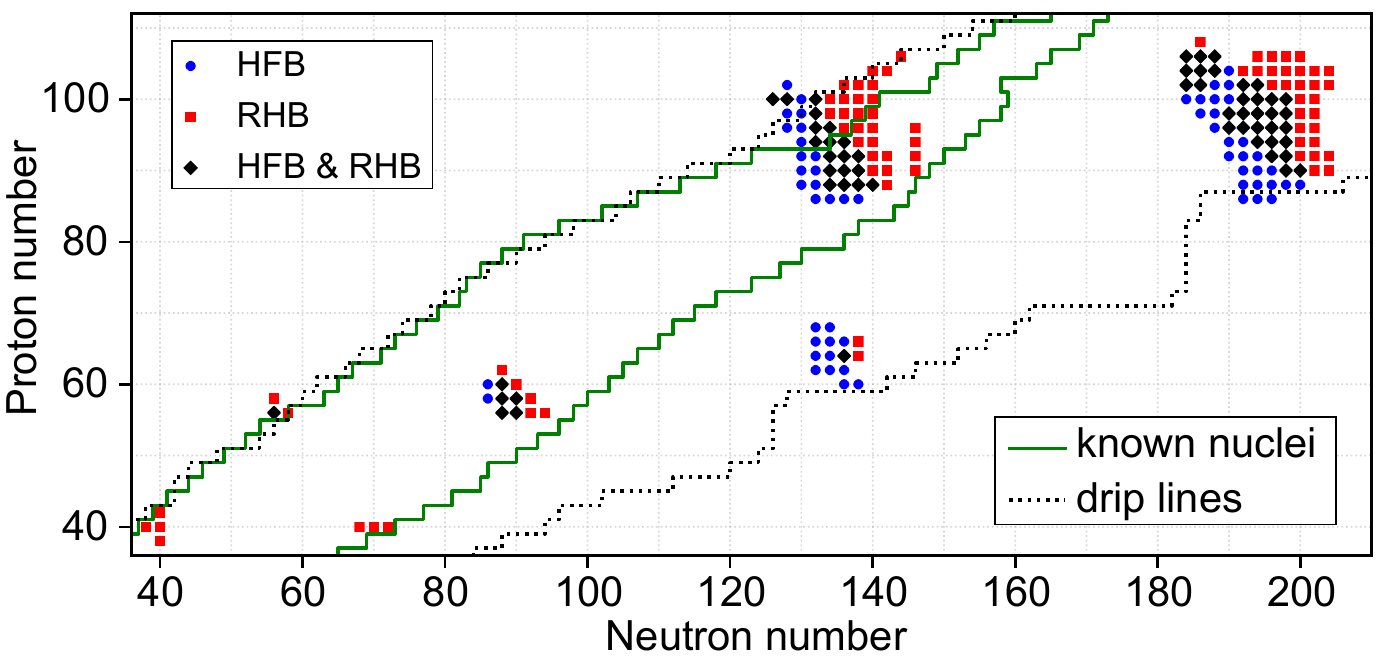}
\caption{\label{cedf_shift}
Comparison between HFB and RHB predictions. Dots mark the HFB predictions with $m\ge 3$, squares show   the RHB predictions with $m\ge 2$, and diamonds mark the overlap region between HFB and RHB results.
The borders of known nuclei and two-particle drip lines are 
 as in  Fig.~\ref{multiplicity}.
}
\end{figure*}

Apart from the overall agreement  between SEDFs and CEDFs when it comes to the predicted regions of octupole-instability, we see systematic shifts (by 2-4 neutrons) between the regions of $\Delta E_{\rm oct}$ and $\beta_3$ obtained by
 these two  EDF families.  This systematic effect is illustrated  in Fig.~\ref{cedf_shift}, where  dots mark the HFB predictions with $m\ge 3$, squares show   the RHB predictions with $m\ge 2$, and diamonds mark the overlap of the two. 
This shift has been noticed in Ref.~\cite{Agbemava2017} pertaining to superheavy nuclei.

\section{Local trends}\label{local}

The majority of octupole-deformed nuclei are found near the intersection between neutron numbers 88, 134, and 194 and proton numbers 56 and 88. This pattern is more pronounced in heavy nuclei, due to their  lower values of $\Delta e$, see Fig.~\ref{sp_split}. 

We note that all of the EDFs used in this study  provide robust and consistent predictions for quadrupole moments, which generally agree well with available experimental data~\cite{Gaffney2013,Gulda1998,Singh2011}, see \cite{SM}. This suggests that the quadrupole collectivity is well developed. On the other hand, 
in many nuclei, the  octupole deformation energy   has a modest value of less than 500 keV.
Such small values of $\Delta E_{\rm oct}$ indicate soft PESs resulting in  an  octupole collectivity of transitional character, i.e.,  between octupole rotational and vibrational motions \cite{Butler1996}. While in this work  we refer to a nucleus as octupole-deformed when its g.s. has  $\beta_3\ne 0$, this does not mean that this octupole deformation is static.  For  octupole-soft, transitional nuclei,  beyond mean-field methods are needed to describe  the system \cite{Egido1991,Robledo2010,Robledo2012,Robledo2015,Robledo2016,Yao2015,Yao2016,Xia2017}.

  In the following, we discuss the local regions of  octupole collectivity with  a focus on the cases robustly predicted to be octupole-deformed
 in Skyrme EDF calculations. A detailed discussion of CEDF results can be found in Refs.~\cite{Agbemava2016,Agbemava2017}.
Note that this discussion is not intended to provide a comprehensive review. For a detailed experimental discussion and other recent 
calculations we refer the reader to Refs.~\cite{Butler2016} and \cite{Moller2008,Robledo2011,Xia2017,Ebata2017}, 
respectively.

\begin{figure*}[!htb]
\includegraphics[width=0.8\linewidth]{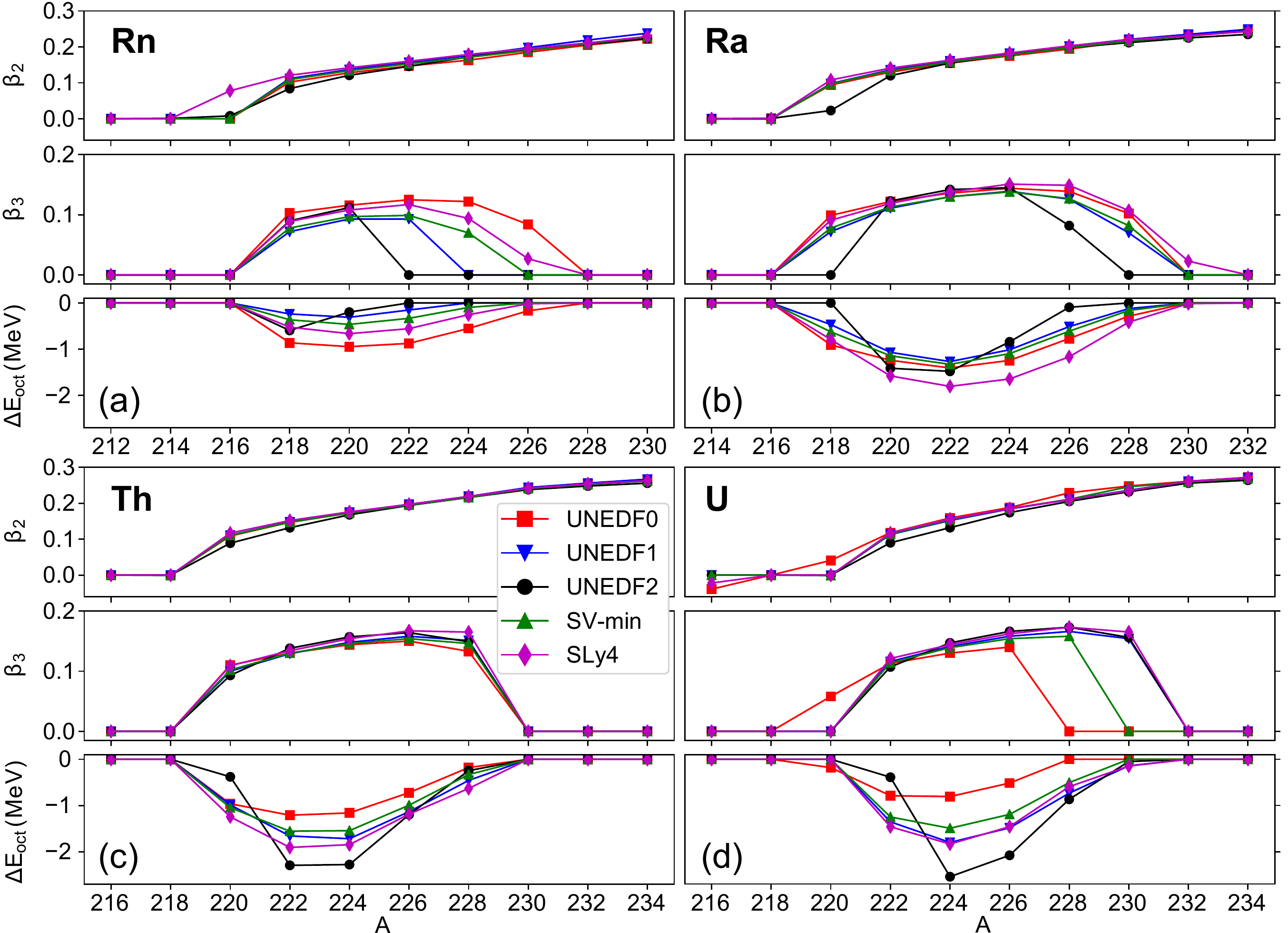}
\caption{\label{actinides}  Values of $\beta_2$, $\beta_3$, and $\Delta E_{\rm oct}$ predicted by the SEDF models  for the isotopic chains of Rn (a), Ra (b), Th (c), and U (d).
}
\end{figure*}

\subsection{Neutron-deficient actinides}

Because of large octupole correlation effects and experimental accessibility, neutron-deficient actinides have traditionally been in the spotlight of octupole deformation studies. As seen in Fig.~\ref{multiplicity}, 
this region is abundant in octupole-deformed nuclei, with many systems predicted robustly by several models, i.e., having  high octupole multiplicity.
Figure~\ref{actinides} summarizes our SEDF results  for the isotopic chains of Rn, Ra, Th, and U.

The isotopes $^{218,220}$Rn and $^{224,226}$Rn have been found experimentally to be close to the octupole vibrational limit~\cite{Gaffney2013,Cocks1997,Cocks1999,Butler2019}.
 As seen in Fig.~\ref{actinides}(a),  $|\Delta E_{\rm oct}|$  reaches its maximum  for  $^{220}$Rn, with an average value around $-0.5$\, MeV. These shallow octupole minima suggest that neutron-deficient Rn isotopes are transitional systems, consistent with experiment. 

The search for octupole instability in neutron-deficient Ra isotopes has been of great interest~\cite{Butler2016,Gaffney2013,Butler2020}, also because of atomic EDM studies. According to numerous theoretical calculations, $^{224}$Ra has the largest octupole deformation~\cite{Butler2016,Butler2020}, and is often predicted to have the largest $\Delta E_{\rm oct}$ among the Ra isotopes. It is therefore hardly surprising that $^{224}$Ra, along with $^{226}$Ra, is predicted to be octuple-deformed by all nine EDFs studied.

Within the SEDFs, the values of  $\beta_2$, $\beta_3$, and $\Delta E_{\rm oct}$ appear to be very consistent for $^{220,224}$Ra, cf. Fig.~\ref{actinides}(b). The largest $|\Delta E_{\rm oct}|$ is predicted for  $^{222}$Ra, followed by $^{220}$Ra and $^{224}$Ra. Recent experiments suggest $^{222}$Ra has the largest octupole deformation among the Ra isotopes followed by $^{226}$Ra, $^{228}$Ra, and  $^{224}$Ra~\cite{Butler2020}. 

Experimentally, even-even
$^{222-226}$Th exhibit many signatures of stable octupole deformation~\cite{Ackermann1993,Smith1995,Cocks1997}, in agreement with  the SEDFs' predictions shown in Fig.~\ref{actinides}(c).  All SEDFs predict octupole deformations in $^{220,222,224,226,228}$Th.

The majority of SEDFs predict even-even $^{222-228}$U to be octupole-deformed. As seen in  Fig.~\ref{actinides}(d), the largest octupole deformation energy exceeding 2\,MeV is calculated for $^{224}$U, followed by $^{222,226}$U. Experimentally, the nucleus $^{226}$U has octupole characteristics similar to  $^{222}$Ra and $^{224}$Th~\cite{Greenlees1998}. According to our study, the nucleus  $^{224}$U  is a superb candidate for a pear-shaped system.

Neutron-deficient Pu isotopes have received little attention in octupole-instability studies as they are extremely difficult to access. The lightest-known Pu isotope,  $^{228}$Pu, has a half-life of 1.1\,s \cite{Nishio2003} but spectroscopic information
about this system is nonexistent. Likewise, 
virtually nothing is known about $^{230,232,234}$Pu, except for their g.s. properties~\cite{NNDC}. Interestingly the isotope  $^{228}$Pu is predicted by all our models to be octupole-deformed, followed by $^{226}$Pu ($m=7$) and $^{230}$Pu ($m=8$).
The large values of  $|\Delta E_{\rm oct}|$ in $^{224,226,228}$Pu (1.5-2 MeV) calculated by SEDFs  are similar to those Ra, Th, and U isotopes that show evidence for  stable octupole deformations. 

The lightest Cm isotope known experimentally is $^{233}$Cm, which is significantly heavier than our best Cm candidates for pear-like shapes: $^{228,230}$Cm. 
As seen in Fig.~\ref{multiplicity}, in neutron-deficient actinides with $Z \geq 98$, most of the best candidates for octupole deformation  lie well beyond the current discovery range, and some appear to be close, or outside, the predicted two-proton drip line 
~\cite{landscape2020}. 

\subsection{Neutron-rich lanthanides}

\begin{figure}[!htb]
\includegraphics[width=0.8\linewidth]{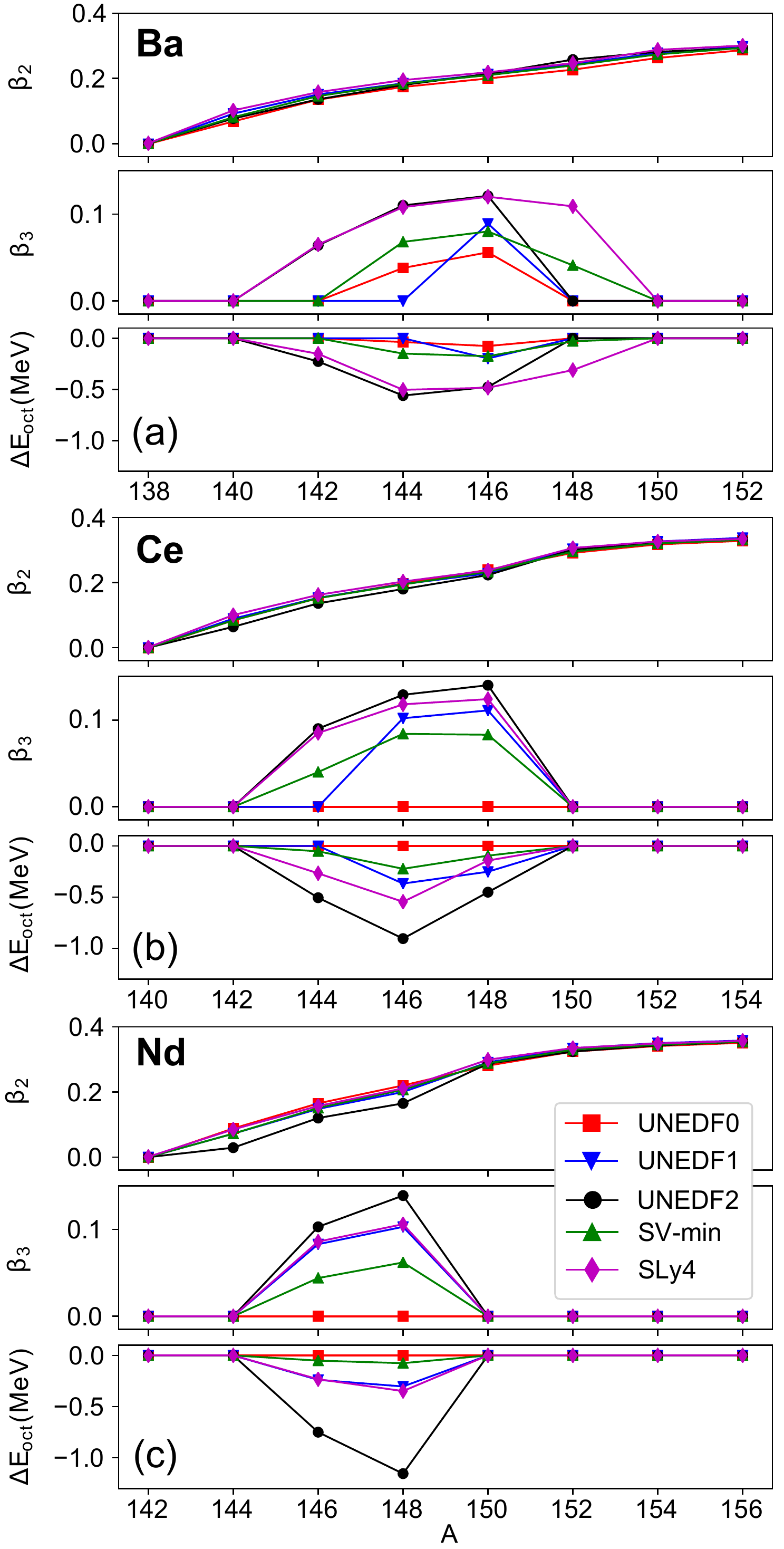}
\caption{\label{lanthanides} Similar  to  Fig.~\ref{actinides} but for Ba (a), Ce (b), and Nd (c).
}
\end{figure}

The region of  Ba, Ce, Nd, and Sm isotopes  around $^{146}$Ba constitutes
 the second largest concentration of octupole-unstable nuclei predicted theoretically that are within experimental range.  Figure~\ref{lanthanides} summarizes our SEDF results  for the isotopic chains of Ba, Ce, and Nd.

Intrinsic dipole moment measurements indicate appreciable octupole correlations in even-even $^{140-148}$Ba~\cite{Phillips1986,Mach1990,Zhu1995,Urban1997}. In particular, direct measurements of E3 transition strength made recently in $^{144,146}$Ba~\cite{144Ba,146Ba} suggest  similar octupole correlations in these nuclei (within large experimental uncertainties). As seen in Fig.~\ref{lanthanides}(a), except for UNEDF1, the SEDF  results are consistent with this  discovery by predicting similar  $\beta_3$ and $\Delta E_{\rm oct}$ for these isotopes.

For the Ce isotopes, all SEDFs except UNEDF0 predict octupole deformations in $^{146,148}$Ce, with the largest $|\Delta E_{\rm oct}|$ in $^{146}$Ce, see Fig.~\ref{lanthanides}(b).  Experiment suggests enhanced octupole correlations in $^{146}$Ce~\cite{146Ce}, $^{144}$Ce~\cite{146Ce,Zhu1995}, and $^{148}$Ce~\cite{Chen2006}, and a weakened octupole collectivity in  $^{150}$Ce \cite{Zhu2012}.

The stable isotopes $^{146,148}$Nd are predicted to be octupole-deformed.  Experimental data suggests enhanced octupole collectivity in $^{146,148,150}$Nd \cite{Urban1981,Iacob1996,Ibbotson1993,Ibbotson1997,Friedrichs1992}.
Another stable isotope with high octupole multiplicity is $^{150}$Sm.  Experimentally, there is some evidence for octupole collectivity in an excited band of $^{150}$Sm \cite{Bvumbi2013}. As seen in Fig.~\ref{multiplicity}, the  isotopes 
$^{146,148,150}$Nd and $^{150}$Sm are the only stable even-even candidates for  octupole instability. The parity doublets in odd-mass nuclei from this region, such as $^{153}$Eu,
can be excellent candidates for searches of T,P-violating effects
 with atoms, ions, and molecules \cite{Flambaum2020}.

\subsection{Proton-rich nuclei around $^{112}$Ba}

Strong octupole correlations, including octupole instability, were predicted theoretically in nuclei around $^{112}$Ba in the early 1990s \cite{Skalski1990,Heenen1994}. 
 As seen in  Fig.~\ref{multiplicity}, some of our models yield reflection-asymmetric shapes  in a handful of nuclei from this region 
that  lie close to, or beyond, the proton drip-line, with $^{112}$Ba being the best candidate.
The experimental data in this region are scarce,   with enhanced octupole correlations being  suggested for $^{112}$Xe~\cite{Smith2001} and $^{114}$Xe \cite{Rugari1993}.
The lightest observed Ba isotope is $^{114}$Ba \cite{Capponi2016}, for which no spectroscopic information exists.

\subsection{Proton-rich and neutron-rich zirconium regions} 

Shallow octupole minima are calculated in the Zr region around $N=40$ and $N=70$ by some CEDFs, see Fig.~\ref{cedf_shift}, and also by Gogny calculations of Ref.~\cite{Robledo2011}. On the other hand, our SEDF models predict no octupole instability in this region.

\subsection{Very neutron-rich  nuclei around $^{200}$Gd}

Many extremely neutron-rich nuclei around $^{200}$Gd are predicted to be octupole-deformed,  see Fig.~\ref{multiplicity} and Supplemental Material \cite{SM}.
While this region lies well outside experimental reach, nucleosynthesis calculations
suggest that it can be accessed  in a very neutron-rich $r$ process \cite{Lippuner2015}.
The best candidates for octupole instability in this region are $^{196,198,200}$Sm, $^{196,198,200}$Gd, and $^{200}$Dy. 

\subsection{Very neutron-rich  nuclei around $^{288}$Pu}

Many extremely neutron-rich actinide and transactinide nuclei with $184 <N<206$  are predicted to be pear-shaped, see Fig.~\ref{multiplicity} and Refs.~\cite{Erler2012a,Warda2012,Xu2017,Agbemava2017}. 
From a purely nuclear structure perspective, this broad region of octupole instability  is of solely theoretical interest as it lies well outside experimental reach. While the production of nuclei heavier than $N=184$ in the astrophysical $r$ process 
is expected to be strongly hindered by  neutron-induced fission \cite{Giuliani2018a,Giuliani2019}, the magnitude of this suppression  strongly depends on predicted fission barriers 
\cite{Giuliani2020} and hence  the question of their astrophysical relevance is still open.

\section{Summary}\label{summary}

  A systematic survey of reflection-asymmetric axially symmetric even-even nuclei has been 
carried out within the Skyrme-HFB approach. 
Among the five SEDFs employed, UNEDF2 and SLy4 predict the largest number of octupole-deformed nuclei, and also the largest octupole deformation energies $\Delta E_{\rm oct}$. The functional UNEDF0, which was not optimized to experimental shell gaps, predicts the lowest number of  octupole minima. This can be attributed to the larger energy splitting $\Delta e$ between octupole-driving $(\ell,j)$ and $(\ell-3,j-3)$ shells in this model. The low values of $\Delta e$ are
indicative of enhanced octupole correlations. However, these quantities are not instrumental 
in the predictions of their magnitude.

 These results are combined  with those obtained with four CEDFs in Refs.~\cite{Agbemava2016,Agbemava2017}
and additional RHB calculations performed for the present manuscript. This makes
it possible to produce the landscape of octupole deformations
shown in Fig. ~\ref{multiplicity} that displays reflection-asymmetric nuclei  for non-relativistic and relativistic  EDFs, thus limiting systematic model  uncertainties.
 There are 12 even-even nuclei predicted by all nine EDFs to be octupole-deformed:  $^{146}$Ba, $^{224,226}$Ra, $^{226,228}$Th, $^{228}$Pu, $^{288,290}$Pu, $^{288,290}$Cm, and $^{288,290}$Cf.

By analyzing the  trend of $\beta_2$, $\beta_3$, and $\Delta E_{\rm oct}$ along isotopic chains of actinides and lanthanides, we found that the SEDF results are fairly consistent with other  studies \cite{Moller2008,Robledo2011,Robledo2012}. The study of Ref.~\cite{Ebata2017}  predicted relatively few octupole-unstable nuclei, which was acknowledged by the authors as possibly due to the strong pairing interaction. A shift in the position of octupole-unstable regions (by 2-4 neutron numbers) is  seen when comparing the results of SEDF and CEDF models. This shift can be seen in Fig.~\ref{cedf_shift}, and likely comes from the shell structure obtained with CEDF models, as the SEDF results agree well with the results of other global  non-relativistic surveys. 
Minor differences aside, the octupole landscape presented in Fig.~\ref{multiplicity}, is consistent with current experimental data.
 Quadrupole deformations are highly consistent across all nine models used and  agree well with experiment.

In the neutron-deficient actinide region, in addition to the ``usual suspects", our study suggests stable g.s. octupole deformations in  $^{224,226,228}$U,$^{226,228,230}$Pu, and $^{228,230}$Cm. The only stable pear-shaped even-even nuclei expected theoretically are $^{146,148,150}$Nd and $^{150}$Sm.

Our global survey predicts two exotic regions of octupole instability in extremely neutron-rich nuclei that are inaccessible experimentally. The first region, of lanthanide nuclei  around $^{200}$Gd, is possibly populated in a very neutron-rich $r$ process. In the   second region of actinide and transactinide nuclei with $184 <N<206$, neutron-induced fission is likely to suppress the $r$-process production of  
nuclei with  $N>184$, but the magnitude of this hindrance strongly depends on predicted fission barriers.

It will be of great interest to expand  this work by systematic DFT studies of octupole deformation and underlying single-particle structure
in  odd-mass and odd-odd nuclei. Progress has been made in exploring  particle-odd systems by using projection techniques, primarily in the
systematic computation of Schiff moments \cite{Jacek2018}, but much work still remains to be done.

\begin{acknowledgments}
Discussions with Jacek Dobaczewski and Samuel Giuliani are greatly appreciated. Computational resources were provided by the Institute for Cyber-Enabled Research at Michigan State University. This material is 
based upon work  supported by the U.S.\ Department of Energy, Office of Science, Office of Nuclear Physics under  award numbers DE-SC0013365 (Michigan State University) 
and DE-SC0013037 (Mississippi State University), DE-SC0018083 (NUCLEI SciDAC-4 collaboration),   DOE-NA0003885 and DOE-NA0002925
(NNSA, the Stewardship Science Academic
Alliances program), and DE-SC0015376 (DOE Topical Collaboration in Nuclear Theory for Double-Beta Decay and Fundamental Symmetries).
\end{acknowledgments}


%

\end{document}